\documentclass[twocolumn]{aastex631}

\usepackage[T1]{fontenc}
\usepackage{ae,aecompl}

\usepackage{graphicx}	% Including figure files
\usepackage{amsmath}	% Advanced maths commands
\usepackage{amssymb}	% Extra maths symbols
\usepackage{booktabs,tabularx}
\usepackage{enumitem}
\usepackage{multirow}
\usepackage{hyperref}

\def\gax{\mathrel{\raise.3ex\hbox{$>$}\mkern-14mu\lower0.6ex\hbox{$\sim$}}}
\def\lax{\mathrel{\raise.3ex\hbox{$<$}\mkern-14mu\lower0.6ex\hbox{$\sim$}}}
\def\gtorder{\mathrel{\raise.3ex\hbox{$>$}\mkern-14mu
             \lower0.6ex\hbox{$\sim$}}}
\def\ltorder{\mathrel{\raise.3ex\hbox{$<$}\mkern-14mu
             \lower0.6ex\hbox{$\sim$}}}

\begin{document}

\title[The Mid-IR Counterpart to N6946-BH1]{The search for failed supernovae with the Large Binocular
  Telescope: The Mid-IR Counterpart to N6946-BH1}

\shortauthors{Kochanek, Neustadt, \& Stanek}

%%%%%%%%%%%%%%%%%%%%%%%%%%%%%%%%%%%%%%%%%%%%%%%%%%%%%%%%%%%%%%%%%%%%%%%%%

\author{Christopher S. Kochanek}\thanks{Email: kochanek.1@osu.edu}
\affiliation{Department of Astronomy, The Ohio State University, 140 W. 18th Ave., Columbus, OH 43210, USA}
\affiliation{Center for Cosmology and AstroParticle Physics, The Ohio State University, 191 West Woodruff Ave., Columbus, OH 43210, USA}

\author{Jack M. M. Neustadt}
\affiliation{Department of Astronomy, The Ohio State University, 140 W. 18th Ave., Columbus, OH 43210, USA}

\author[0000-0001-7351-2531]{Krzysztof Z. Stanek}
\affiliation{Department of Astronomy, The Ohio State University, 140 W. 18th Ave., Columbus, OH 43210, USA}
\affiliation{Center for Cosmology and AstroParticle Physics, The Ohio State University, 191 West Woodruff Ave., Columbus, OH 43210, USA}

%%%%%%%%%%%%%%%%%%%%%%%%%%%%%%%%%%%%%%%%%%%%%%%%%%%%%%%%%%%%%%%%%%%%%%%%%

% Abstract of the paper
\begin{abstract}
We present JWST MIRI 5.6, 10 and $21\mu$m observations of the candidate failed supernova N6946-BH1
along with HST WFPC/IR 1.1 and $1.6\micron$ data and ongoing optical monitoring data with the LBT.
There is a very red, dusty source at the location of the candidate which has only $\sim 10$-$15$\% of the luminosity of the progenitor star.  The source is very faint in
the HST near-IR observations ($\sim 10^3 L_\odot$) and is not optically variable to a limit of
$\sim 10^3 L_\odot$ at R-band. The dust is likely silicate and probably has to be dominated
by very large grains as predicted for dust formed in a failed SN.  The required visual
optical depths are modest, so we should begin to have a direct view of the source in the
near-IR in only a few years.
\end{abstract}

%%%%%%%%%%%%%%%%%%%%%%%%%%%%%%%%%%%%%%%%%%%%%%%%%%%%%%%%%%%%%%%%%%%%%%%%%

%%%%%%%%%%%%%%%%%%%%%%%%%%%%%%%%%%%%%%%%%%%%%%%%%%%%%%%%%%%%%%%%%%%%%%%%%
\section{Introduction}\label{sec:intro}

The formation rate of stellar mass black holes (BH) is a critical unknown for the physics of
massive stars, supernovae (SNe), nucleosynthesis, the origin of X-ray (and other) binaries, and gravitational
wave sources.  In modern theories of massive star death, stellar mass BHs form in failed
SN without a high luminosity explosion.  Stars either explode to form a neutron star, or
they do not, and form a BH (e.g., \citealt{OConnor2011}, \citealt{Ugliano2012}, \citealt{Ertl2016},
\citealt{Sukhbold2016}).  These models suggest that 10--30\% of core collapses lead to
a failed SN.  This is supported by the lack of higher mass SN progenitors (\citealt{Kochanek2008},
\citealt{Smartt2009}, but see \citealt{Davies2020} and \citealt{Kochanek2020}).

For red supergiants (RSGs), a failed supernova should have a weak transient
in which most of the hydrogen envelope is ejected to leave a BH with roughly the mass of
the helium core (e.g., \citealt{Lovegrove2013}, \citealt{Lovegrove2017}, \citealt{Fernandez2018},
\citealt{Ivanov2021}, \citealt{Antoni2023}), with significant dust formation in the ejected envelope (\citealt{Kochanek2014b}).  The ejection of
the envelope in a failed SN of an RSG provides the first natural explanation of the
$5$--$10M_\odot$ BHs found in the galaxy (\citealt{Kochanek2014a}, \citealt{Kochanek2015}).

Measuring the rate of BH formation and failed SNe is challenging because the most
unambiguous signatures would be neutrino or gravitational wave observations of the
collapse, and these will only be feasible in our Milky Way galaxy and its satellites
for the foreseeable future.  But if the Galactic SN rate is roughly one per century,
there is a failed SN only once every 300-1000~years (see, e.g., \citealt{Adams2013}).

In \cite{Kochanek2008}, we proposed the one presently feasible approach to finding failed
SNe. With a 8m class telescope, we can search for RSGs which
vanish in a failed SN independent of the nature of any intermediate transient.  We
have been carrying out this program for 15 years (\citealt{Gerke2015}, \citealt{Adams2017a},
\citealt{Neustadt2021}), finding one very good candidate (\citealt{Gerke2015})
and one weaker candidate (\citealt{Neustadt2021}).  If we just count the strong candidate, this implies a
failed SN rate of $f_{SN}=0.16_{-0.12}^{+0.23}$ at 90\% confidence that is consistent
with theoretical expectations (\citealt{Neustadt2021}).  It is reassuring that the number
of candidates is small, because it means that there are few or no other sources that
mimic the vanishing of an RSG in a failed SN.

The best candidate, named
N6946-BH1 for its host galaxy and status as the first candidate, had a weak
transient in 2009 and then rapidly faded to become optically invisible to levels
$< 1\%$ of the pre-transient luminosity.  Prior to the transient, the progenitor had a nearly constant
optical luminosity that could be traced back to the late 1990s.
When \cite{Basinger2021} (and previously \citealt{Adams2017b}) last did a detailed study,
they found 
no optical counterpart to limits of $\ltorder 10^3L_\odot$ or less, a low-luminosity
fading near-IR counterpart ($\sim 2000 L_\odot$) and no mid-IR counterpart
($\ltorder 10^4 L_\odot$) compared to an estimated progenitor luminosity of
$\simeq 10^{5.6}L_\odot$. 

\cite{Basinger2021} were left with two issues. First, the Spitzer mid-IR constraints
were weaker than desirable, subject to problems from confusion due to their
low angular resolution, and could only constrain the presence of relatively
warm dust emission since they provided measurements only at $3.6$ and $4.5\mu$m.
While obscuration of a star by an ongoing dusty wind and some models of
obscuration by dust formed in the 2009 transient could be ruled out, the
limits could be evaded by simply making the dust more distant and colder at
the price of pushing upwards the ejecta mass and kinetic energies into a
physically challenging regime if the present day luminosity is the same as
the progenitor luminosity.  
Second, if a BH was formed, it should have an extended period of accretion,
and very little mass fall-back is required to produce significant emission
since the Eddington limit for a $10M_\odot$ black hole is $10^{5.5}L_\odot$
(coincidentally matching the progenitor luminosity) and only
requires an accretion rate of $\dot{M}=2 \times 10^{-7}M_\odot$/year for
an efficiency of 10\%. However, the accretion rates are not
well-determined, particularly in the scenario of a failed SN (e.g., \citealt{Perna2014},
\citealt{Quataert2019}, \citealt{Antoni2022}, \citealt{Antoni2023}).

Here we report on new JWST MIRI mid-IR observations, HST WFPC3/IR near-IR
observations and continuing LBT optical monitoring observations of NGC~6946~BH1.
We describe the data and its reduction in \S2.  \cite{Beasor2023} reported
results of their independent JWST observations of the same source taken one month
earlier. We focus on modelling our data and compare to their results where
appropriate.
We model the resulting spectral
energy distribution (SED) and discuss the implications in \S3, and summarize
the results in \S4.  We adopt the same distance, 7.7~Mpc (\citealt{Anand2018}),
and extinction, $E(B-V)=0.303$ (\citealt{Schlafly2011}), used by \cite{Basinger2021}.
The distance is consitent with the $7.8$~Mpc distance found using the
TRGB by \cite{Murphy2018}.

\section{Data}\label{sec:methods}

\begin{figure}
\centering
\includegraphics[width=0.50\textwidth]{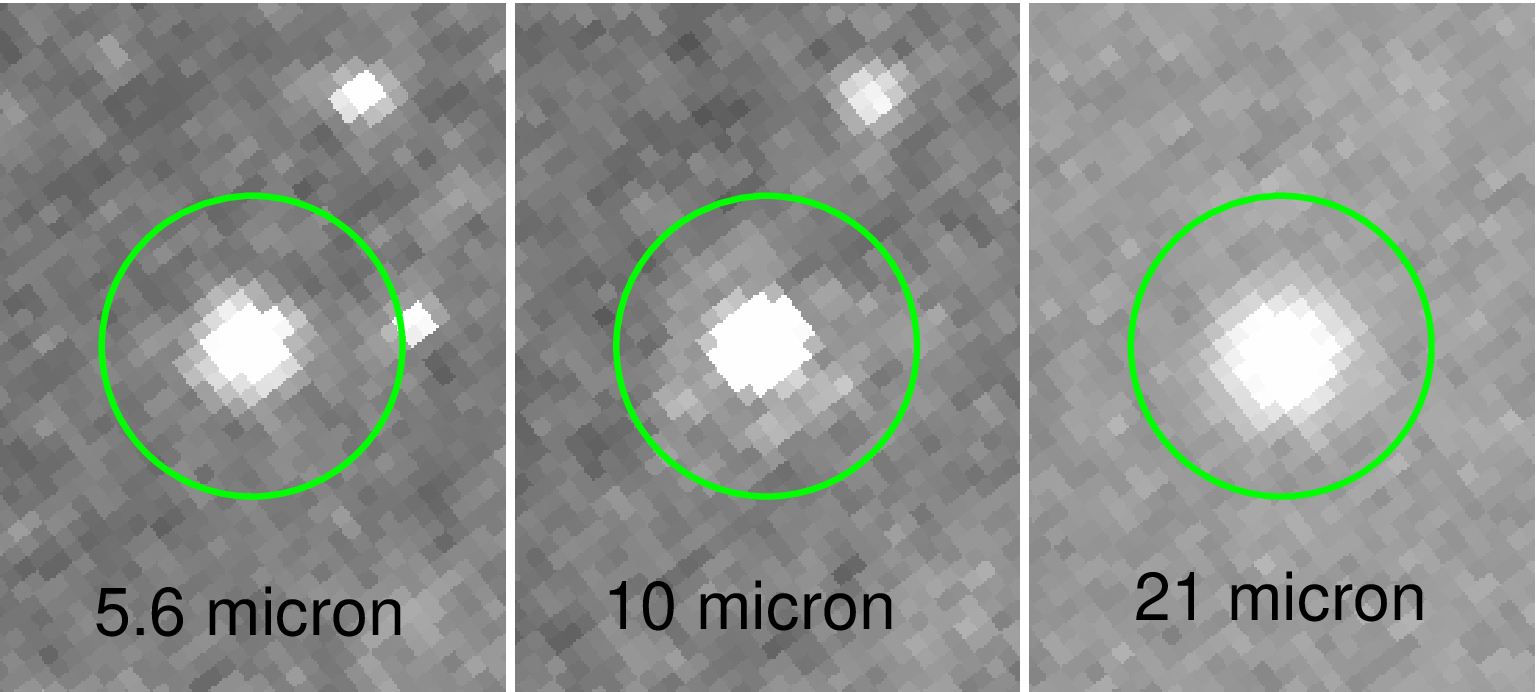}
\caption{
  The JWST MIRI F560W (left), F1000W (middle) and F2100W (right)
  images of N6946-BH1.  The continued brightness of BH1 towards
  longer wavelengths compared to the nearby stars demonstrates that
  the emission is due to dust.
  North is up and the circles are 1\farcs0 in radius.
  }
\label{fig:miri}
\end{figure}

\begin{figure}
\centering
\includegraphics[width=0.50\textwidth]{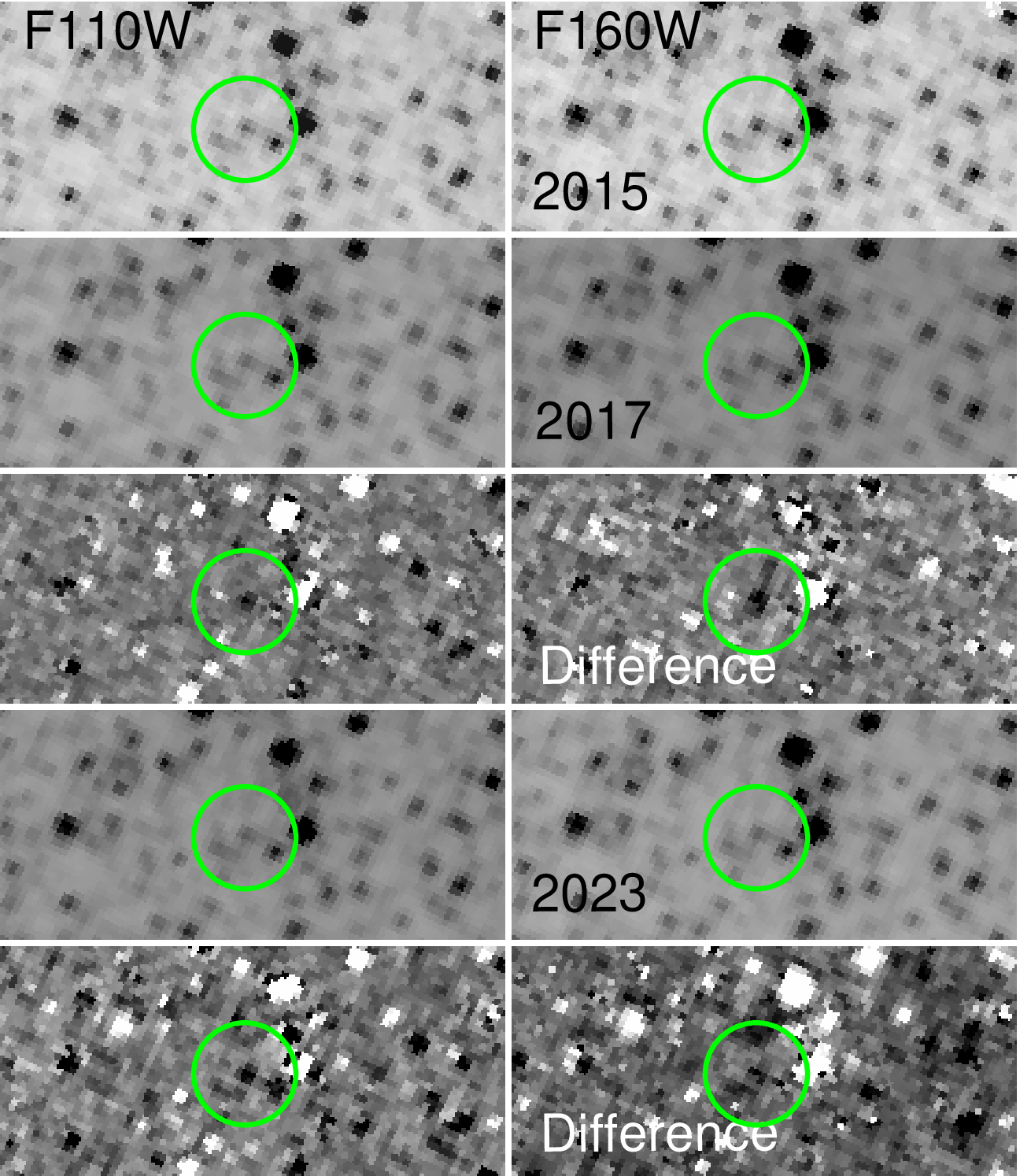}
\caption{
  The HST F110W (left) and F160W (right) observations from 2015
  (top), 2017 (middle) and 2023 (bottom) along with the difference
  images between the 2017/2023 epochs and the 2015 epoch.  Black
  in the difference images means the source has faded relative to
  2015.  North
  is up and the circles are 1\farcs0 in radius centered on
  NGC~1646-BH1.
  }
\label{fig:hst}
\end{figure}

\begin{figure*}
\centering
\includegraphics[width=\textwidth]{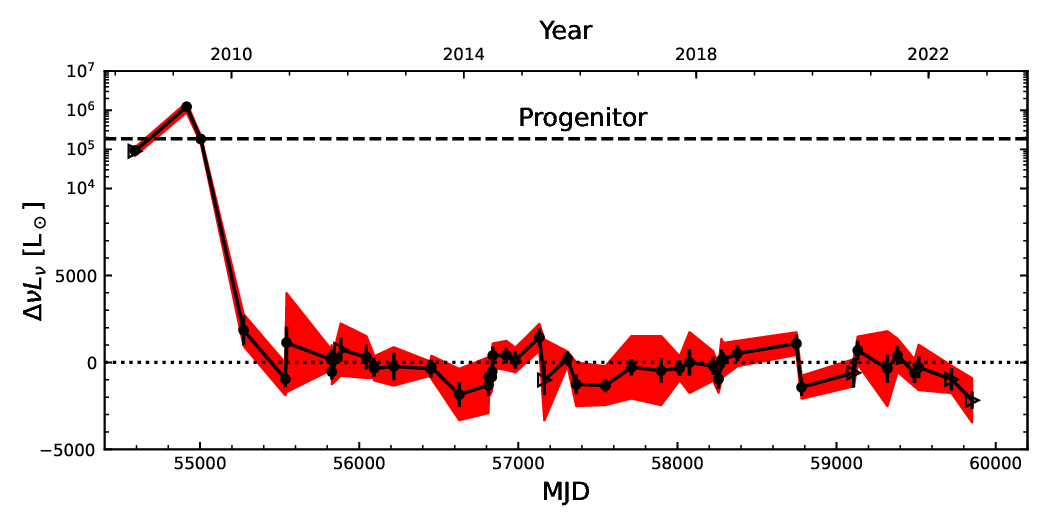}
\caption{
  The LBT R-band light curve of N6946-BH1.  After the optical transient ends in 2010 there have been no optical changes at the level of a few  $10^3L_\odot$ or roughly 1\% of the progenitor luminosity.  Triangles are
  epochs with bad seeing or lower transparency (clouds/cirrus).
  }
\label{fig:lbt}
\end{figure*}

We obtained HST WFPC3/IR F110W (J) and F160W (H) images on 2023 Jan 29
with 3 dithered images in each band and total integration times of
1350~sec for both bands.  We obtained JWST MIRI F560W ($5.6\mu$m),
F1000W ($10\mu$m) and F2100W ($21\mu$m) images on 2023 September 26
with effective exposure times of 87, 87 and 526~seconds. We used
the BRIGHTSKY region with 4 dither positions.  For F560W and F1000W
we used one integration with 25 groups for each image, while for
F2100W we used 3 integrations each with 50 groups. We also have
an ongoing optical monitoring program with the Large Binocular
Telescope (LBT).  The JWST images
are show in Fig.~\ref{fig:miri}, the sequence of HST near-IR
images we have obtained are shown in Fig.~\ref{fig:hst}, and
the LBT R band light curve is shown in Fig.~\ref{fig:lbt}.

The mid-IR source is well-isolated, so we used the STScI pipeline
photometry for the fluxes.  These show small variations ($\sim 10\%$)
between apertures which are unimportant given that we are interested
in the logarithmic value of the luminosity. We did do our own
aperture and PSF photometry as a check of the pipeline values, finding
good agreement.  We obtained fluxes of $17.6\pm 1.8$, $40.6\pm0.3$
and $87\pm 1$ microJanskys for the F560W, F1000W and F2100W bannds,
respectively.  The F560W flux is $\sim 50\%$ of the \cite{Beasor2023}
value, while the other two fluxes agree well.
We used difference imaging (\citealt{Alard1998}, \citealt{Alard2000})
to analyze the new HST data, as this
provides excellent sensitivity to changes in brightness even in the
very crowded near-IR environment of BH1.  We estimate current
F110W and F160W magnitudes of $24.34$ and $22.61$ (dominated by
systematic errors) where the source faded slightly at F110W and
brightened slightly at F160W compared to 2017.
The LBT data is processed using standard methods as described in \cite{Gerke2015},
\cite{Adams2017a} and \cite{Neustadt2021} and
analyzed using difference imaging following \cite{Alard1998} and
\cite{Alard2000}.  We find that the optical flux cannot have changed
by more than $\sim 10^3L_\odot$ at R band since the transient faded
in 2010-2011.

We will fit our MIRI results and use our new HST near-IR results and
the earlier HST optical results from \cite{Adams2017b} as upper limits.  \cite{Beasor2023} find
that the HST resolution near-IR source seems to be three similar luminosity
sources at JWST resolution. This is consistent with the difference between our HST/F160W 
luminosity estimates and their NIRCAM/F185W estimates, but our HST/F110W
luminosities match their NIRCAM/F115W estimates.  Since they also argue that
their NIRCAM/F250M fluxes must be off by a factor of $\sim 2.6$, we decided to just
focus on our data.  We are mainly concerned with the bulk energetics and the
exact values of these near-IR fluxes are not important to the basic result, although
they are crucial to the future evolution of the source as we discuss in \S4.

\begin{figure}
\centering
\includegraphics[width=0.5\textwidth]{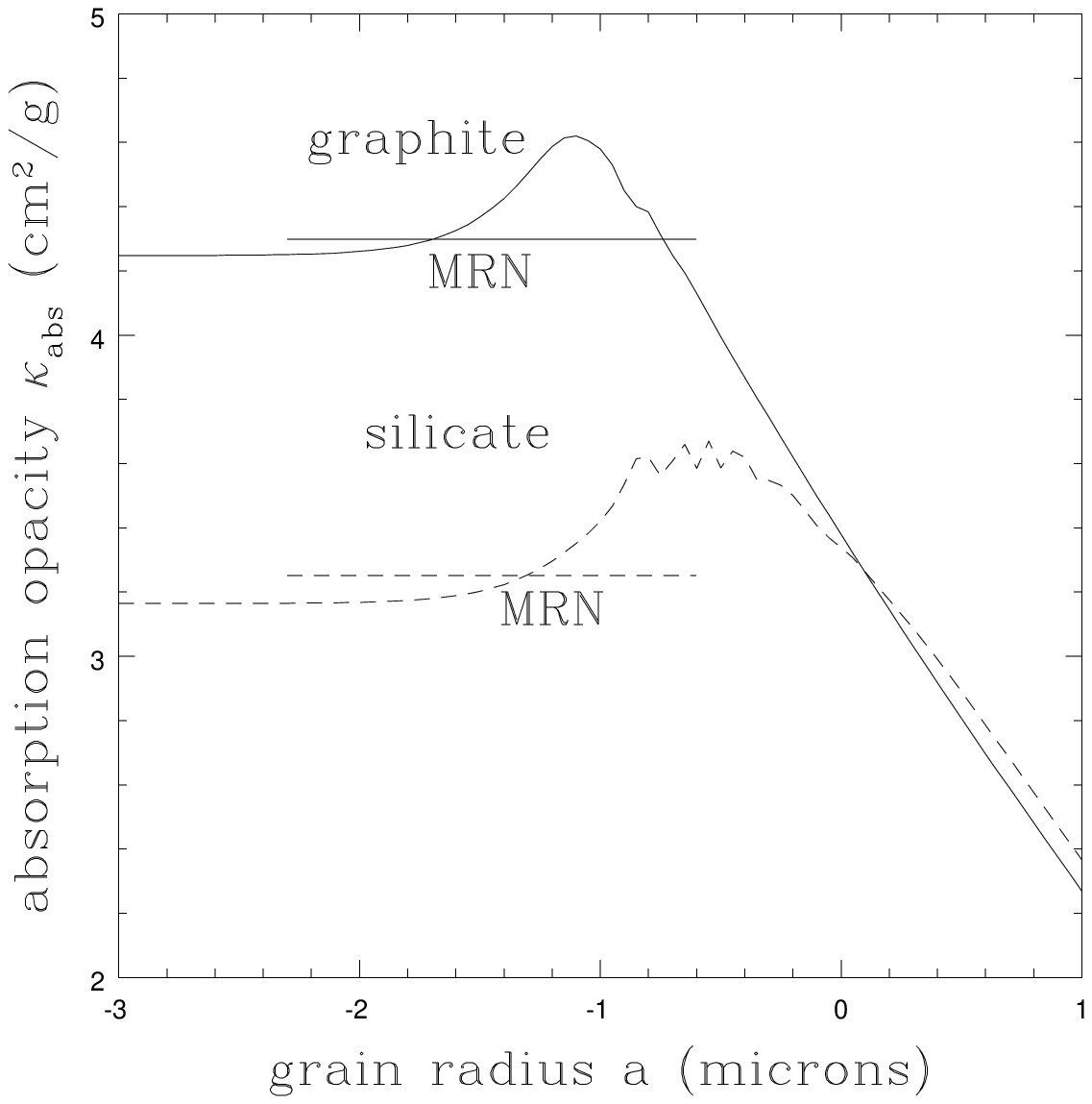}
\caption{
  The V band ($0.55\mu$m) absorption opacities, $\kappa_{abs}$, for
  \protect\cite{Draine1984} graphitic (solid) and silicate (dashed) dusts.  The horizontal lines show the 
  mean for the defualt {\tt DUSTY} \protect\cite{Mathis1977} $a^{-3.5}$ size distribution with $0.005 < a < 0.25\mu$m (the size range spanned by the line).  
  }
\label{fig:dust}
\end{figure}

\section{Results}\label{sec:results}

Following our approach in \cite{Adams2017b} and \cite{Basinger2021}, we model
the spectral energy distribution with {\tt DUSTY} (\citealt{Ivezic1997},
\citealt{Ivezic1999}, \citealt{Elitzur2001}), a spherically symmetric 
dust radiation transfer program.  We model the underlying source using a
\cite{Castelli2003} model stellar atmosphere with a luminosity of $L_*$
and a temperature of $T_*$.  The dust is distributed as a $\rho \propto r^{-2}$
shell from $R_{in}$ to $R_{out}=2 R_{in}$.  In our previous studies we have
found that the results are insensitive to the choice of $R_{out}$ because 
the optical depth is dominated by the inner regions.  We use \cite{Draine1984}
graphitic and silicate dusts and start with a \cite{Mathis1977} $a^{-3.5}$ grain size
distribution spanning $0.005 < a < 0.25\mu$m.  The dust parameters are the visual optical depth $\tau_V$
through the shell, and the dust temperature $T_d$ at the inner edge.  The
dust temperature combined with $L_*$ then determines $R_{in}$, from which
we can estimate the ejecta velocity $v_e \simeq R_{in}/\Delta t$, where
we adopt $\Delta t = 14$~years.  Following \cite{Adams2017b}, {\tt DUSTY} is embedded in a 
Markov Chain Monte Carlo (MCMC) ``wrapper'' which is used to optimize the
fits to the SED and estimate uncertainties. For the most part, parameter
uncertainties are not an important part of our story.

Fig.~\ref{fig:dust} shows the V band ($0.55\mu$m) absorption opacities, $\kappa_{abs}$,
per unit dust mass for the \cite{Draine1984} graphitic and silicate dusts along with their
averages for the \cite{Mathis1977} size distribution.  For consistency
with {\tt DUSTY}, we assume a bulk dust density of $3$~g~cm$^{-3}$ and
a gas-to-dust ratio of $f_{gd}=200$.  The opacity per unit gas mass is then
the opacity per unit dust mass shown in Fig.~\ref{fig:dust} divided by $f_{gd}$.
The \cite{Mathis1977} averaged absorption opacities per unit gas mass 
are then $\kappa_V \simeq 98$~cm$^2$/g and $10$~cm$^2$/g,
for the graphitic and silicate dusts, respectively. The ejecta gas
mass is then
\begin{eqnarray}
     M_{\rm e} &= &{ 4 \pi R_{\rm in} R_{\rm out} \tau_V \over \kappa_V} \\
       &= &0.63 \tau_V { R_{\rm out} \over R_{\rm in} } \left( { R_{\rm in} \over 10^{17}~\hbox{cm}} \right)^2
          \left( { 100~\hbox{cm}^2/\hbox{g} \over \kappa_V } \right) M_\odot \nonumber
\end{eqnarray}
and the ejecta kinetic energy is
\begin{eqnarray}
     K_{\rm e} &= &{ 2 \pi R_{\rm in}^3 R_{\rm out} \tau_V \over \Delta t^2 \kappa_V} \\
       &= &0.032\tau_V { R_{\rm out} \over R_{\rm in} } \left( { R_{\rm in} \over 10^{17}~\hbox{cm}} \right)^4
          \left( { 100~\hbox{cm}^2/\hbox{g} \over \kappa_V } \right) \hbox{FOE} \nonumber
\end{eqnarray}
For the estimated of $M_{\rm e}$ and $K_{\rm e}$ we just use $R_{out}=R_{in}$ since we have
no real constraints on $R_{out}$.  This means that the estimates are lower bounnds on the
mass and the kinetic energy.

\begin{figure}
\centering
\includegraphics[width=0.5\textwidth]{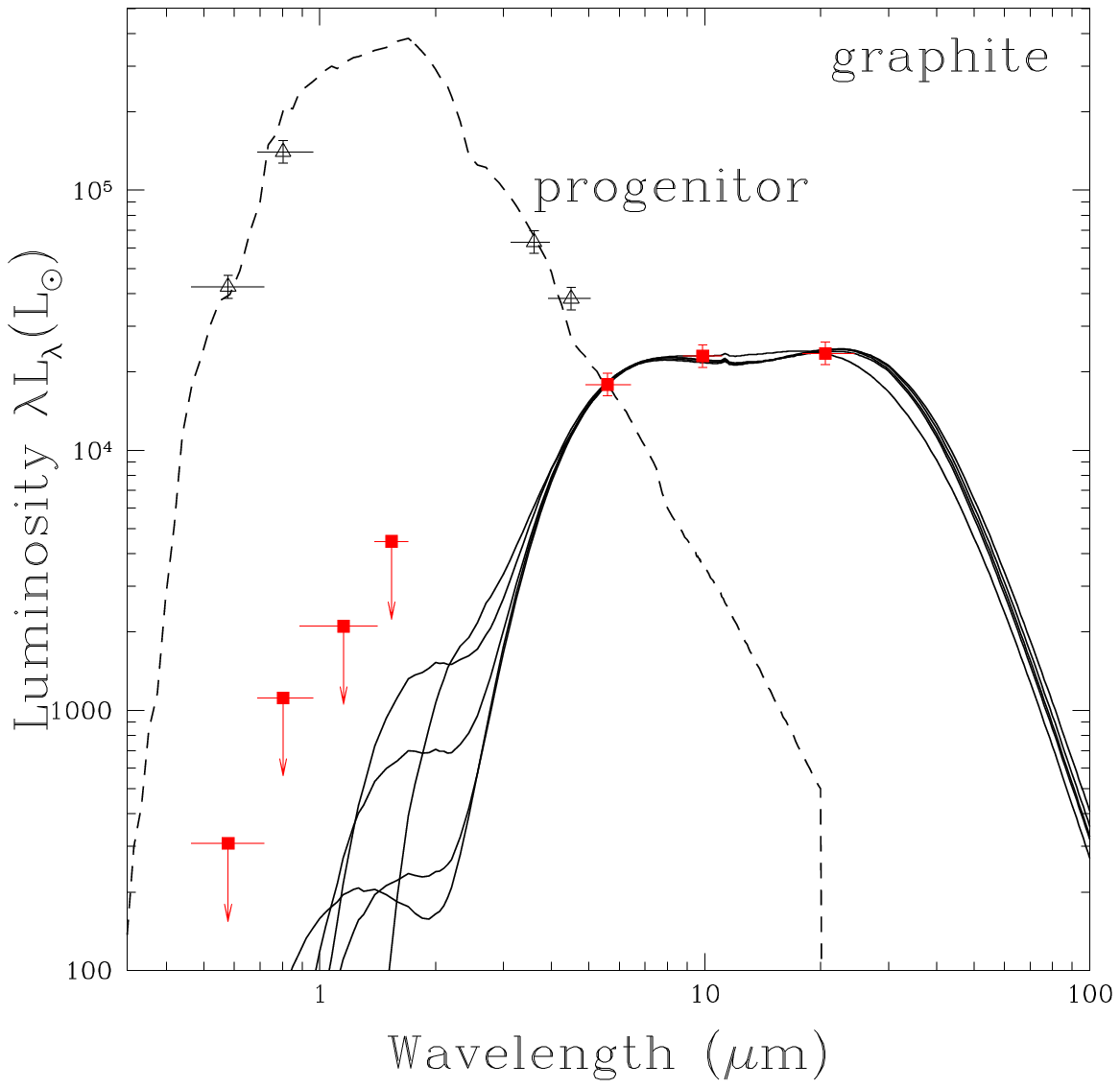}
\caption{
  The present day SED of NGC~6946 (filled red points and limits) modeled with graphitic dust (solid curves)
  for stellar temperatures
  of $T_*=4000$, $T_*=7000$, $T_*=10000$, $T_*=15000$ and $T_*=20000$~K.  The progenitor SED and its model
  are shown by the open triangles and dashed curve.  The horizontal bars on the points are the filter widths.
  }
\label{fig:graphite}
\end{figure}

\begin{figure}
\centering
\includegraphics[width=0.5\textwidth]{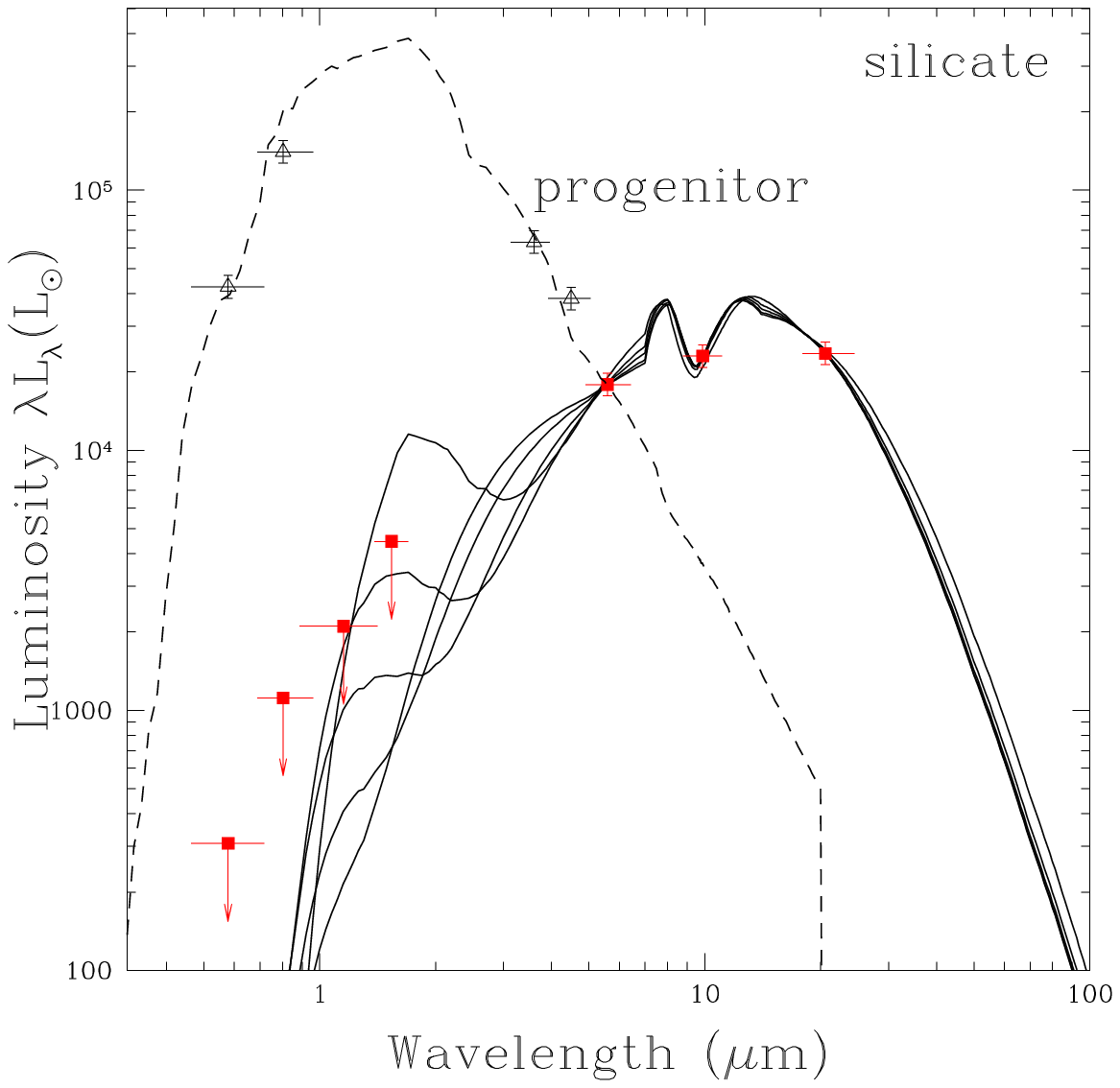}
\caption{
  The present day SED of NGC~6946 (filled red points and limits) modeled with silicate dust (solid curves)
  for stellar temperatures
  of $T_*=4000$, $T_*=7000$, $T_*=10000$, $T_*=15000$ and $T_*=20000$~K.  The progenitor SED and its model
  are shown by the open triangles and dashed curve.  The horizontal bars on the points are the filter widths.
  }
\label{fig:silicate}
\end{figure}

\begin{figure}
\centering
\includegraphics[width=0.5\textwidth]{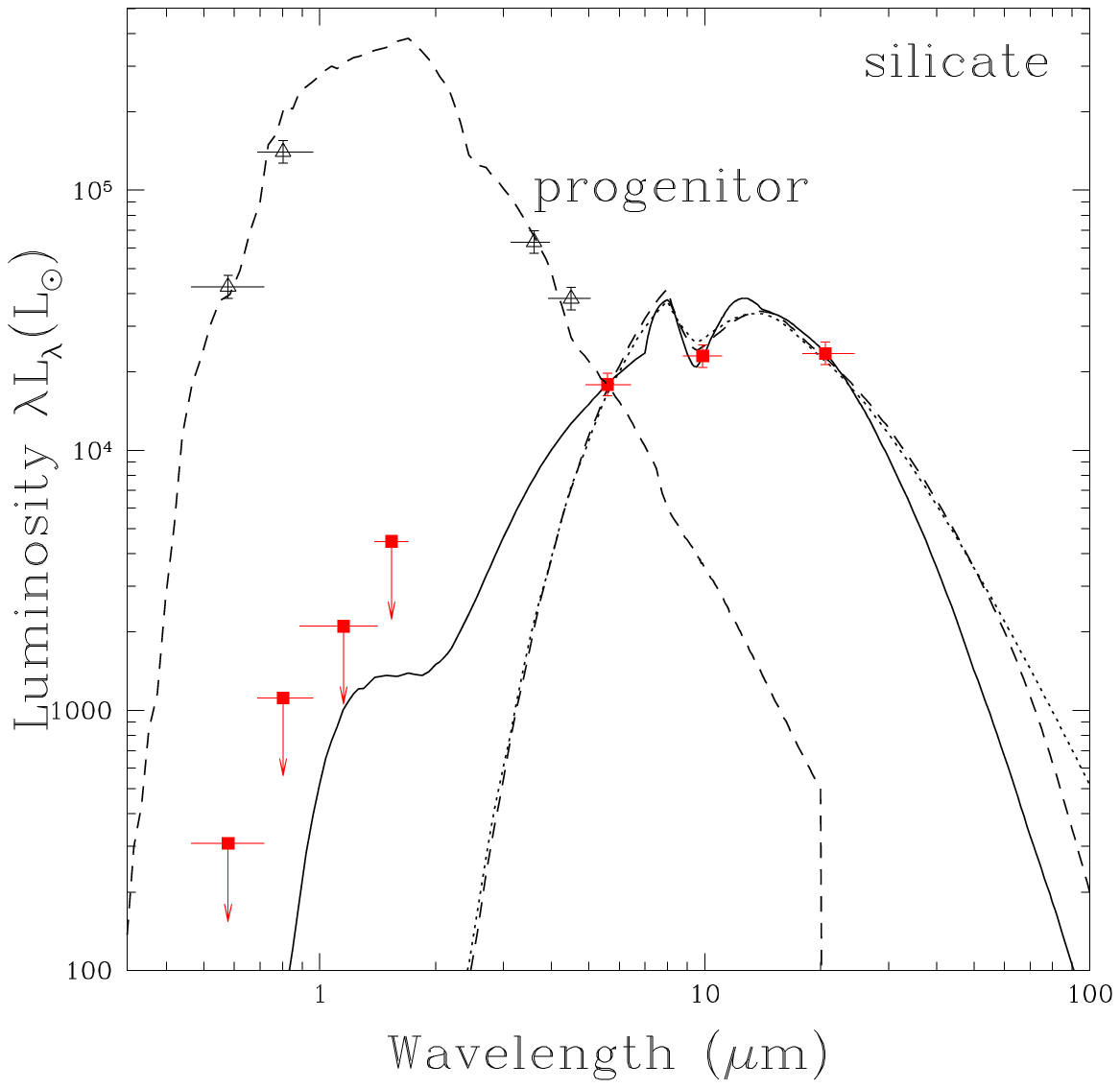}
\caption{
  The present day SED of NGC~6946 (filled red points and limits) modeled with silicate dust (solid curves) for a stellar temperature of $T_*=10000$~K and the grain distribution stopping at
  $a_{max}=0.25\mu$m (solid), $10\mu$m (dashed) and $1000\mu$m (dotted).
    The progenitor SED and its model
  are shown by the open triangles and dashed curve.  The horizontal bars on the points are the filter widths.
  }
\label{fig:biggrain}
\end{figure}

We fit the SEDs assuming stellar temperatures of $T_*=4000$, $7000$, $10000$, $15000$ and
$20000$~K.  With no information on the structure of the optical SED, there is no way to
determine $T_*$.   We fit the SED treating the
optical (F606W and F814W) and near-IR (F110W and F160W) HST luminosities as upper limits
and vary $T_d$, $\tau_V$ and $L_*$ to fit the MIRI luminosities. We include no additional
host extinction.  The primary effect of adding host extinction is that it drives up the
estimate of the progenitor luminosity while having a negligible effect on the mid-IR
luminosity observed today.  This would exacerbate the differences between the present
day and progenitor luminosities.  

The key comparison for these models is between the luminosity
implied by the current SED and that of the progenitor. Here we 
compare to the progenitor properties found using MARCS (\citealt{Gustafsson2008})
stellar atmospheres fit to the archival HST F606W and F814W data from 2007 along
with the nearly concurrent Spitzer 3.6 and $4.5\mu$m images (see \citealt{Adams2017b}).  
We use the MARCS models
here because the preferred temperatures lie close to the lowest 
temperatures in  the \cite{Castelli2003} models.  Models with no dust fit resonably
well with $T_*\simeq 3500$~K and $L_*=10^{5.63}L_\odot$. The photometric constraints
and the progenitor SED model are shown in Figs.~\ref{fig:graphite} and \ref{fig:silicate}.
Models with a little
hot dust ($T_d\simeq1300$~K, $\tau_V\simeq 0.2$) fit a little better, but the stellar 
parameters change little ($T_* \simeq 3500$~K, $L_*-10^{5.62}L_\odot$).  The higher luminosity
here and in \cite{Basinger2021} is driven by using a larger distance (7.7 versus 6.0~Mpc)
than \cite{Adams2017b}. 

These initial models are really focused on the possibility of a surviving star surrounded
by ``normal'' dust.  The primary difference from \cite{Adams2017b} and \cite{Basinger2021}
is that the mid-IR emission now constrains the actual present day luminosity.  Figs.~\ref{fig:graphite}
and \ref{fig:silicate} show the SEDs and their models for graphitic and silicate dust,
respectively, for all five source temperatures $T_*$. The models are able to
fit the mid-IR fluxes while staying below the optical/near-IR limits for
both dust types and all stellar temperatures.  As expected, the input spectrum
largely only matters in the near-IR, where the limits force any contributon to the overall
luminosity to be modest.  The mid-IR SEDs varies only a little bit with $T_*$ and this
is mostly due to the freedom created by the luminosity uncertainties. The
mid-IR SEDs differ between the dust types because the silicate dusts have 
a feature near $10\mu$m while the graphitic dusts do not.  The graphitic dusts
have a flat SED at peak, while the silicate dusts produce a double peaked SED with
the $10\mu$m data lying in the valley between the peaks.  The silicate models predict that
a peak should be seen at $7\mu$m in the F770W band, and this is exactly what 
\cite{Beasor2023} observed.  This indicates that the dusts are likely the
silicate dusts expected for massive RSGs.

The model luminosities are almost exactly the same with a range of $10^{4.70}L_\odot$
to $10^{4.80}L_\odot$.  This means that the present day luminosity is only 10-15\% of
the progenitor luminosity, and the survival of the progenitor is essentially ruled out.
The visual optical depths are functions of both the dust type and $T_*$, with optical
depths for graphitic dust and higher temperatures.  But at least for these models,
they are not extreme, with $\tau_V=6.9$ to $38$ for the graphitic dusts and
$19$ to $33$ for the silicate dusts.  The dust temperatures would be better
constrained with measurements at $2$ to $3\mu$m, but range from a minimum of 
$T_d \simeq 420$ to a maximum of $1000$~K.  The dust radius depends on the stellar
temperature because the Planck absorption factors $Q(T_*)$ are larger for hotter 
stars, forcing the dust to be more distant for higher stellar temperatures.  For
the graphitic dusts, the radii are in the range $R_{in}=10^{15.8}$ to $10^{16.3}$~cm
and for the silicate dusts they are all roughly $10^{15.5}$~cm.   This implies 
relatively slow expansion velocities of $150$-$500$~km/s for the graphitic models
and $\sim 80$~km/s for the silicate models.
Because the optical depths are modest and the low luminosity allows $R_{in}$ to be
relatively small, the necessary ejecta masses are small, $M_e \sim 0.1$ to $0.2M_\odot$
which, combined with the low velocities, means the kinetic energies are also small
compared to SN energies, $\sim 10^{-5}$ to $5 \times 10^{-4}$~FOE.  

The low masses may seem problematic for a failed SN interpretation even if the 
velocities are of the right order of magnitude.  Given the near-IR detections,
the problem cannot be solved by simply increasing the optical depth, nor is
preventing dust formation a likely solution (see \S4). The solution has to lie
in the opacity, in particular from using the
\cite{Mathis1977} grain size distribution.  In the \cite{Kochanek2014b} failed
SN dust formation models, there are very few grains in the \cite{Mathis1977}
size range -- 98\% (!) of the dust mass is in grains bigger than $10\mu$m.
Since the large grains have optical Planck factors $Q \simeq 1$, their
absorption cross sections are simply $\propto a^2$ while their masses are
$\propto a^3$, so all of the dust mass is in big grains while the absorption
is still dominated by smaller grains.
If we simply extend the maximum grain size
the silicate absorption opacities per unit gas mass drop from $\kappa_{abs}\simeq 10$
to $5.4$, $1.9$ and $0.6$~cm$^2$/g for $a_{max}=10$, $100$ and $1000\mu$m.  Fig.~2
in  \cite{Kochanek2014b} shows that only the outermost, lowest density layers of
the ejecta produce grains smaller than $1\mu$m.  With an additional small change
in the size distribution from \cite{Mathis1977} in favor of larger grains, it
appears likely that the visual opacities can be driven down far enough to allow
ejecta masses of many $M_\odot$.  As seen in Fig.~\ref{fig:biggrain}, extending
the grain size distributions to very large values of $a_{max}$ has little effect
on the SED model.

\section{Discussion}\label{sec:discussion}

While we make no use of the \cite{Beasor2023} data, we note that essentially
all of our silicate dust models predict enhanced emission at $7.7\mu$m compared
to 5.6 and $10\mu$m, which is what they found.  The usual logic for dust formation
is that the type of dust is determined by the abundance of carbon relative to
oxygen.  The two elements will preferentially bind to make CO, and then 
graphitic dusts are made if there is left over carbon and silicate dusts
are made if there is left over oxygen.  In the MIST stellar models
(\citealt{mist1}, \citealt{mist2}, \citealt{mist3}, \citealt{mist4}, \citealt{mist5})
the envelopes of RSGs are all oxygen rich.  
RSG winds are also observed to be
dominated by oxide dusts like the silicates (e.g., \citealt{Verhoelst2009})
and the same would be expected for
dust formed in the ejected envelope since the winds simply have the 
composition of the envelope.  They need not be simply silicate dusts,
but exploring additional types is beyond our present scope.

As discussed in \cite{Adams2017b}, the optical luminosity of the progenitor
was fairly steady over the decade prior to the transient in 2009.  \cite{Adams2017b}
fit SEDs including Spitzer mid-infrared contstraints and found no significant
change in the bolometric luminosity between July 2005 and July 2008, although
there may have been some modest changes in a dusty wind.  So there is no good
evidence for the pre-merger luminosity variations seen in V1309~Sco (\citealt{Tylenda2011}).
The brightness changes in V1309~Sco were also quite modest, about 1~mag at I-band, with most
of that occurring in the last $~\sim 3$~years.  In our case, the R band optical luminosity
was steady for the last decade particularly compared to the order of magnitude
difference between the luminosity of the progenitor and the luminosity today. The energetically
less important B and V bands varied more, fading in the last few years
combined with a rise in the mid-IR
emission, consistent with a change in the wind optical depth but at a nearly 
constant bolometric luminosity.  It is, of course, presently popular to invoke
mass loss changes in RSGs shortly before core collapse (e.g., \citealt{Smith2023}
for SN~2023ixf recently).

\cite{Kashi2017} argue that obscuring the system with a dusty disk viewed edge
on would allow most of the luminosity to escape in the polar directions and thus
significantly reduce the total observed flux, thereby breaking the link between
the observed luminosity and the luminosity of the underlying source.  \cite{Beasor2023}
also invoke this concept as part of their argument for making this system a 
stellar merger.  However, \cite{Adams2017b} had already demonstrated for the
limiting case of a slab geometry that this idea does not work in practice.
More recently, \cite{Kochanek2023} carried out a detailed investigation of dust radiation 
transfer in disky geometries and found that even for extremely high disk
optical depths (up to $\tau_V=10^3$, orders of magnitude larger than needed here), the luminosity inferred by an observer in the disk plane was
at most a factor of 2 different from the true luminosity.  It is also 
difficult to suppress the optical flux by the extreme factors needed here
(see Figs.~\ref{fig:lbt}, \ref{fig:graphite}, \ref{fig:silicate}),
because even a small amount of polar dust will scatter a detectable
amount of optical light to the observer.  

V1309~Sco largely demonstrates this point because the progenitor was an
eclipsing binary, and hence our viewing angle is close to the plane of the
orbit (\citealt{Tylenda2011}).  While the optical emission dropped to be fainter
than the progenitor in roughly two years, the total luminosity remained 
$\sim 60$ times greater than the pre-event luminosity ($\sim 9 L_\odot$) four years after peak
(\citealt{Tylenda2016}).  By 2017, an epoch roughly comparable to our present
observations, its I, J and K$_s$ band luminosities were $3.2$, $2.5$, and $0.4L_\odot$
(\citealt{Ferreira2019}), so its total luminosity in just these three bands 
approaches the total luminosity of the progenitor.  V838~Mon also shows
a disk like structure  (\citealt{Chesneau2014}, \citealt{Kaminski2021},
\citealt{Mobeen2021}) but never became significantly fainter than the progenitor 
and it presently has a similar B band flux to the progenitor and is brighter
at V band (compare \citealt{Munari2005} and \citealt{Liimets2023}). This is
complicated by the presence of a triple companion, but it is argued that the
star producing the transient was the more luminous.  Based on \cite{Munari2005},
the progenitor luminosity would be $\sim 10^4 L_\odot$ and the present day
luminosity including the dust emission  (e.g., \citealt{Woodward2021})
appears to be comparable or higher.  

  \cite{Beasor2023} express
concerns that the X-ray  and ultraviolet radiation produced by accretion would
destroy the dust.  This is an issue when grains are first forming because high
energy photons can stochastically heat small grains above the evaporation 
temperature and suppress dust formation (\citealt{Kochanek2011}).  For an
existing grain, the energy
of an ionizing photon is shared over many atoms/bonds because it is deposited
by a fast electron Coulomb scattering through the grain.  The electron loses
its energy over a distance $\simeq 0.01 E_{keV}^{1.5}\mu$m for a photon
of energy $E_{keV}$~keV (\citealt{Draine1979}), which is nearly $10^6 E_{keV}^{1.5}$ 
atomic spacings -- the energy deposited per atom/bond is actually quite small.
 Grain shattering by 
electrically charging the grains due to interactions with X-rays may
be feasible, but requires the very high fluences of $\gamma$-ray bursts
(see \citealt{Waxman2000}, \citealt{Fruchter2001}). Moreover,
even if the accretion luminosity
commenced immediately, the sheer amount of ejected mass protects most of the
ejecta during the dust formation phase.  For example, if we have material expanding
at 200~km/s, one solar mass of hydrogen has a column density of $2\times 10^{26} t_{year}^{-2}$~cm$^{-2}$,
which is Compton thick for over a decade (see the discussion of the X-ray 
absorption in \citealt{Basinger2021}).  Similarly, the inner layers of gas
would absorb all the ionizing ultraviolet flux for an extended period of time.
In short, the inner gas protects the outer gas while it forms dust and once the dust is formed
it basically does not care about the energy of the photons.

Thus, we argue that the simplest explanation remains interpreting N6946-BH1
as a failed SN currently powered by accretion luminosity.  But it is a hypothesis
that should certainly undergo additional tests.  It would be nice to find 
additional pre-event observations hiding in various archives, but this seems
unlikely at this point.  This really only leaves continued monitoring of the 
source, where these observations and those of \cite{Beasor2023} make it clear that
this needs to be done with JWST.  While we have no constraints from our observations
on the NIRCAM filters, we do find a significantly different ($\sim 50\%$) flux in our MIRI
F560W fluxes (we agree on the $10$ and $21\mu$m fluxes). Variability seems to be a 
natural consequence of accretion, so a JWST monitoring campaign should be informative. 
Densely filling in the SED with all the available NIRCAM/MIRI filters should clarify
the nature of the dust, and at least in the mid-IR a spectrum could likely be obtained in
a reasonable integration time.
But in the end, time may be the only fundamental test -- either we fade to black, or
we do not. The optical depths we infer are not tremendously high, and mass conservation
requires the dust optical depth to drop at least as fast as $\tau \propto t^{-2}$ --
so even on a one year time scale, a visual optical depth of $\tau_V=20$ should drop
to $\tau_V \simeq 17.5$ and the corresponding K-band optical depth of $\tau_K \simeq 2$
would drop to $\tau_K=1.7$ -- any underlying source should start to become direcly
visible in the near-IR fairly shortly.

\facilities{HST, JWST, LBT}

%%%%%%%%%%%%%%%%%%%%%%%%%%%%%%%%%%%%%%%%%%%%%%%%%%%%%%%%%%%%%%%%%%%%%%%%%

\section*{Acknowledgements}

 JN thanks A. Ash and J. Roberts for help with the MIST models.
 JN and CSK are supported by NSF grants AST-2307385
 and AST-1908570. 
 Support for programs JWST GO-2896 and HST GO-17144 was provided by NASA through a grant from the Space Telescope Science Institute, which is operated by the Association of Universities for Research in Astronomy, Inc., under NASA contract NAS 5-03127.
 This work is based [in part] on observations made with the NASA/ESA/CSA James Webb Space Telescope. The data were obtained from the Mikulski Archive for Space Telescopes at the Space Telescope Science Institute, which is operated by the Association of Universities for Research in Astronomy, Inc., under NASA contract NAS 5-03127 for JWST. These observations are associated with program GO-2896.
This research is based on observations made with the NASA/ESA Hubble Space Telescope obtained from the Space Telescope Science Institute, which is operated by the Association of Universities for Research in Astronomy, Inc., under NASA contract NAS 5–26555. These observations are associated with program(s) GO-17144.
 The LBT is an international collaboration among
 institutions in the United States, Italy, and Germany. LBT Corpora-
 tion partners are: The University of Arizona on behalf of the Arizona
 university system; Istituto Nazionale di Astrofisica, Italy; LBT
 Beteiligungsgesellschaft, Germany, representing the Max-Planck
 Society, the Astrophysical Institute Potsdam, and Heidelberg Uni-
 versity; The Ohio State University, and The Research Corporation,
 on behalf of The University of Notre Dame, University of Minnesota
 and University of Virginia.
 
\section*{Data Availability}

The HST and JWST data will be publically available after the standard proprietary
period.  The LBT light curve is available from the authors.

%%%%%%%%%%%%%%%%%%%%%%%%%%%%%%%%%%%%%%%%%%%%%%%%%%%%%%%%%%%%%%%%%%%%%%%%%
\bibliography{bib}{}
\bibliographystyle{aasjournal}
%%%%%%%%%%%%%%%%%%%%%%%%%%%%%%%%%%%%%%%%%%%%%%%%%%%%%%%%%%%%%%%%%%%%%%%%%

%%%%%%%%%%%%%%%%%%%%%%%%%%%%%%%%%%%%%%%%%%%%%%%%%%%%%%%%%%%%%%%%%%%%%%%%%

\end{document}